\documentclass[10pt,letterpaper]{article}
\usepackage{hyperref}
\usepackage{amsmath}
\usepackage{subfigure}
\usepackage{epsfig}

\begin{document}

%\twocolumn[
\title{\bf Moving Targets Virtually Via Composite Optical Transformation}

\author{Wei Xiang Jiang and Tie Jun Cui\footnote{tjcui@seu.edu.cn}\\
{\small\it State Key Laboratory of Millimeter Waves and Institute of Target Characteristics and Identification}\\
{\small\it Department of Radio Engineering, Southeast University,
Nanjing 210096, P. R. China.}}

\date{}
\maketitle

\begin{abstract}

We propose a composite optical transformation to design an illusion
device which can move the image of a target from one place to
another place. Enclosed by such an illusion device, an arbitrary
object located at one place appears to be at another place
virtually. Different from the published shifted-position cloak which
is composed of the left-handed materials with simultaneously
negative permittivity and permeability, the illusion device proposed
in this letter has positive permittivity and permeability. Hence
proposed illusion device could be realized by artificial
metamaterials.

\hskip 1.0mm


\end{abstract}
%]

\newpage

In the past few years, great attention has been paid to the
transformation optics [1-21]. Among the various optical
transformation devices, the most exciting and fascinating one is the
invisibility cloak [1-7]. The first free-space cloak with simplified
medium parameters was verified experimentally at the microwave
frequency [3]. Recently, a carpet cloak has been proposed [4] and
confirmed in microwave frequencies [5] and then in optical
frequencies [6,7]. Inspired by these pioneering work, other
applications such as the rotation of electromagnetic (EM) waves
[10,11], and omnidirectional retroreflectors [12,13] have been
proposed and investigated.

More recently, Lai \emph{et} al. have proposed an interesting idea
of illusion optics, which makes a target with arbitrary shape and
material properties look like another object of some other shape and
material makeup [19]. Some special illusion devices have been
discussed before this, such as the super absorber [13], the
superscatterer [16], the shifted-position cloak [18], and the
cylindrical superlens [21]. However, all above-mentioned illusion
devices are composed of the left-handed materials with
simultaneously negative permittivity and permeability. Hence such
illusion devices are highly demanding on the material properties,
and the applications are possible to be limited to the realm of
theory.

In this letter, we propose a composite optical transformation to
design an illusion device which makes the enclosed actual object
located at one place appear to be at another place. Such an illusion
device shows unconventional EM properties as verified by accurately
numerical simulations. We try to make the illusion device be fairly
realizable. Unlike the published illusion devices which are composed
of left-handed materials [13, 16-19, 21], all permittivity and
permeability components of the proposed illusion device are
positive. Hence the presented method makes it possible to realize
the illusion device by using the modern metamaterials.

A concise schematic to design the proposed illusion device is
illustrated in Fig. 1. A car at Position $A$ is enclosed with an
illusion device, as shown in Fig. 1(a). Such an illusion device make
any detector outside the virtual boundary (surface $s$) observe the
EM fields of a virtual car at Position $B$ (shown in Fig. 1(c))
instead of the actual one. In other words, the illusion device makes
the EM-field distributions outside the virtual boundary in both the
virtual and physical spaces exactly the same, regardless the
direction from which the EM waves are incident. The composite
coordinate transformation of the illusion device contains two
sub-mappings. First, we map the first virtual space (the free space
outside the car in the virtual space in Fig. 1(c)) into a smaller
annulus region (the light green region in Fig. 1(b)). Second, we
transform the second virtual space (light green region in Fig. 1(b))
into the physical space (dark green region in Fig. 1(a)). The
electric permittivity and magnetic permeability tensors of the
illusion device are calculated by
\begin{eqnarray}
\overline{\varepsilon''}=\mathrm{\Lambda_2}\mathrm{\Lambda_1}\overline{\varepsilon}
\mathrm{\Lambda_1}^{T}\mathrm{\Lambda_2}^{T}/\det(\mathrm{\Lambda_1}\mathrm{\Lambda}_2),~~~~
\overline{\mu''}=\mathrm{\Lambda_2}\mathrm{\Lambda_1}\overline{\mu}
\mathrm{\Lambda_1}^{T}\mathrm{\Lambda_2}^{T}/\det(\mathrm{\Lambda_1}\mathrm{\Lambda_2}),
\end{eqnarray}
in which $(\overline{\varepsilon}$, $\overline{\mu})$ and
$(\overline{\varepsilon''}$, $\overline{\mu''})$ are the
permittivity and permeability tensors in the first virtual space
(illusion space) and the physical space, respectively, and
$\mathrm{\Lambda_1}$ and $\mathrm{\Lambda_2}$ are the Jacobian
transformation matrices with components
$\mathrm{\Lambda_1}_{ij}=\partial x^{'}_{i}/\partial x_{j}$ and
$\mathrm{\Lambda_2}_{ij}=\partial x^{''}_{i}/\partial x^{'}_{j}$
corresponding to the mapping from the first virtual space to the
second virtual space, and the mapping from the second virtual space
to the physical space, respectively.

Similar to other optical transformation devices, the EM fields in
the illusion device can be obtained from the transformation optics
[1,2] as
$\mathbf{E}''=(\mathrm{\Lambda_2}^T)^{-1}(\mathrm{\Lambda_1}^T)^{-1}\mathbf{E}$
and
$\mathbf{H}''=(\mathrm{\Lambda_2}^T)^{-1}(\mathrm{\Lambda_1}^T)^{-1}\mathbf{H}$,
where $\mathbf{E}$ and $\mathbf{H}$ are electric and magnetic fields
in the first virtual space, respectively. Because the virtual
boundary $s$ is mapped to itself during these two sub-mappings
$\mathrm{\Lambda_1}$ and $\mathrm{\Lambda_2}$, we have
$\mathbf{E''}_t=\mathbf{E}'_t=\mathbf{E}_t$ and
$\mathbf{H}''_t=\mathbf{H}'_t=\mathbf{H}_t$, where the subscript $t$
indicates the transverse components along the surface $s$. That is
to say, the tangential components of the EM fields on the whole
virtual boundary ($s$) are exactly the same in both physical and
virtual spaces. Hence the EM fields outside the illusion device are
also exactly the same by using the uniqueness theorem. Any observers
outside the illusion device will perceive the scattered fields as if
they were scattered from a car at Position $B$.

For a specific example, we consider an illusion device which can
move the image of a metallic cylinder from one place to another
place. To design such an illusion device, we construct a composite
transformation which contains two sub-mappings in the Cartesian
coordinate system, the first sub-mapping can be expressed as
\begin{eqnarray}
&&x'=\frac{r_{1}'(x-a_1-a_2)}{r}+a_1+a_2, \hskip 0mm\\
&&y'=\frac{r_{1}'y}{r},\\
&&z'=z,
\end{eqnarray}
where $(a_1+a_2,~0)$, $(0,~0)$ and $(a_1,~0)$ are the centers of
virtual and actual metallic cylinders and circular virtual region,
respectively, $r_{1}'=(R_{12}-R_{11})(r-r_0)/(R_{12}-r_0)+R_{11}$,
$R_{1i}=-(x-a_1-a_2)(a_1+a_2)/r+\sqrt{b_i^2-(a_1+a_2)^2y^2/r^2}$,
$(i=1, 2)$, and $r=\sqrt{(x-a_1-a_2)^2+y^2}$. Such a sub-mapping
describes a similar transformation that maps the blank region in the
virtual space in Fig. 1(c) into the light green region in Fig. 1(b).
The second sub-mapping can be expressed as
\begin{eqnarray}
&&x''=\frac{r''x'}{r_{2}'}, \hskip 32mm\\
&&y''=\frac{r''y'}{r_{2}'},\\
&&z''=z',
\end{eqnarray}
where $r''=(R_{22}-r_1)(r'-R_{21})/(R_{22}-R_{21})+r_1$,
$R_{2i}=-(x'-a_1)a_1/r'+\sqrt{b_i^2-a_1^2y'^2/r'^2}$, ($i=1,2$), and
$r_{2}'=\sqrt{x'^2+y'^2}$. This sub-mapping describes a similar
transformation that maps the light green region in Fig. 1(b) into
the dark green region in Fig. 1(a). The converse transformation of
the virtual variables and the physical variables can be constructed
by Eqs. (2)-(7). Then the constitutive parameters of the illusion
device can be calculated by formulae (1).

We remark that the EM parameters of the illusion device can also be
obtained by another process for such a composite transformation. A
function relationship between $(x'',~y'',~z'')$ and $(x,~y,~z)$ can
be found from Eqs. (2)-(7) as follows,
\begin{eqnarray}
&&x''=f(x,y),\hskip 36mm\\
&&y''=g(x,y),\\
&&z''=z.
\end{eqnarray}
Based on the above composite transformation, the permittivity and
permeability of the illusion device are expressed as
\begin{eqnarray}
&&\varepsilon''_{xx}=\mu''_{xx}=(f_x^2+f_y^2)/(f_xg_y-f_yg_x),\\
&&\varepsilon''_{yy}=\mu''_{yy}=(g_x^2+g_y^2)/(f_xg_y-f_yg_x), \hskip 20mm\\
&&\varepsilon''_{xy}=(f_xg_x+f_yg_y)/(f_xg_y-f_yg_x)=\varepsilon'_{yx},\\
&&\varepsilon''_{zz}=\mu''_{zz}=1/(f_xg_y-f_yg_x),\\
&&\varepsilon''_{xz}=\varepsilon''_{yz}=\mu''_{zx}=\mu''_{zy}=0,\\
&&\mu''_{xy}=\mu''_{yx}=\varepsilon''_{xy},
\end{eqnarray}
where we denote the partial derivative of the function $f(x,y)$ to
$x$ as $f_x$. We have assumed that the background profile be free
space, i.e., $\overline{\varepsilon}=\varepsilon_{0}\overline{I}$
and $\overline{\mu}=\mu_{0}\overline{I}$.

Obviously, these two sub-mappings are not conformal transformations,
hence some non-diagonal components of the permittivity and
permeability tensors are non-zero. But in the real fabrications, it
is necessary that the material parameters $\overline{\varepsilon''}$
and $\overline{\mu''}$ are denoted in diagonal tensors. The symmetry
of the real tensors $\overline{\varepsilon''}$ and
$\overline{\mu''}$ ensure that a rotation transformation which maps
a symmetric tensor into a diagonal one always exists. When a
transverse-electric (TE) polarized plane wave is incident upon an
enclosed metallic cylinder with infinite length, there exists only
$z$ component of electric field, hence only $\mu''_{xx}$,
$\mu''_{xy}$, $\mu''_{yy}$ and $\varepsilon''_{zz}$ are of interest
and must satisfy the request of Eqs. (11)-(16). Therefore, the EM
parameters in the eigen-basis can be expressed as
\begin{eqnarray}
&&\mu_{1}= \frac{\mu_{xx}^{''}+\mu_{yy}^{''}-\sqrt{\mu_{xx}^{''2}-2\mu_{xx}^{''}\mu_{yy}^{''}+\mu_{yy}^{''2}+4\mu_{xy}^{''}}}{2},\\
&&\mu_{2}=\frac{\mu_{xx}^{''}+\mu_{yy}^{''}+\sqrt{\mu_{xx}^{''2}-2\mu_{xx}^{''}\mu_{yy}^{''}+\mu_{yy}^{''2}+4\mu_{xy}^{''}}}{2},\\
&&\varepsilon_{z}=\varepsilon_{zz}^{''},
\end{eqnarray}
and all the off-diagonal components equal zero. In such a case, all
components $\mu_{1}$, $\mu_{2}$, and $\varepsilon_{z}$ are finite
and positive. Also, $\mu_{2}$ is always greater than $\mu_{1}$.

Unlike the earlier-proposed shifted-position cloak which consists of
left-handed materials with double negative constitutive parameters
[18], the illusion device here is only composed of one layer of
inhomogeneous and anisotropic medium. In order to verify the design
of the illusion device, we make accurately numerical simulations
using the software package, COMSOL Multiphysics, which is based on
the finite-element method. We consider an illusion device which
moves a metallic cylinder from one place virtually to another place.
In this example, we choose $a_1=0.1$ m, $a_2=0.05$ m, $r_0=r_1=0.03$
m, $b_1$=0.13 m, and $b_2=0.15$ m.

Under the illumination of TE-polarized waves, only $\mu_1$, $\mu_2$
and $\varepsilon_z$ are of interest. The distributions of principle
components $\mu_1$, $\mu_2$ and $\varepsilon_z$ are illustrated in
Figure 2, from which we clearly observe that all values are finite
and positive. It is worth to note that these two sub-mappings in
this illusion device are compressing and extending mappings,
respectively, instead of the folding of geometry [17-21]. Hence any
material parameters of the illusion device are not negative. Similar
metamaterial structures have been extensively investigated and
fabricated in the experiment of free-space cloaks and other devices
at microwave frequencies [3,11].

Figure 3 illustrates the numerical results of total electric fields
for the above illusion device. The plane waves are incident
vertically from the bottom to the top at 6 GHz. Figures 3(a) and
3(c) show the total-field distributions of the metallic cylinder
located at different places. When the metallic cylinder enclosed by
the shifted-position illusion device, the scattered pattern from the
metallic cylinder will be changed as if the metallic cylinder was at
another place. This can be clearly observed by comparing the
scattered patterns of the metallic cylinder coated by the illusion
device shown in Fig. 3(b) and the metallic cylinder shown in Fig.
3(c). Inside the virtual boundary, the EM-field distributions in
Figs. 3(b) and 3(c) are very different. But the total-field
distributions are exactly the same outside the virtual boundary. A
position-moving illusion has been generated.

In summary, we have presented a composite optical transformation to
design a kind of illusion device, which can move the target from one
place to another place virtually by using metamaterials. In such an
illusion device, all diagonal components of constitutive tensors are
positive and all off-diagonal ones are zero, hence the illusion
device could be realized using artificial metamaterial structures.

%\section*{{{\bf Acknowledgments}}}

This work was supported in part by the National Science Foundation
of China under Grant Nos. 60990320, 60990324, 60671015, 60871016,
and 60901011, in part by the Natural Science Foundation of Jiangsu
Province under Grant No. BK2008031, and in part by the 111 Project
under Grant No. 111-2-05. WXJ acknowledges the support from the
Graduate Innovation Program of Jiangsu Province under No.
CX08B\_074Z.

\newpage

\section*{{{\bf List of Figure Captions}}}

\noindent \textbf{Fig. 1:} {(color online) A concise scheme of an
illusion device that moves the image of a car from one position to
another one. (a) A car at Position $A$ enclosed with the illusion
device (dark green region). (b) A bigger illusion car in the second
virtual space. (c) The illusion car at Position $B$ in first virtual
space.}

\vskip 5mm

\noindent \textbf{Fig. 2:} {(color online) The parameter
distributions of the illusion device, (a) $\varepsilon_z$, (b)
$\mu_1$, and (c) $\mu_2$.}

\vskip 5mm

\noindent \textbf{Fig. 3:} {(color online) The total electric-field
distributions in the computational domain for (a) a metallic
cylinder without the illusion device; (b) the actual metallic
cylinder with the illusion device (dashed circle means a virtual
metallic cylinder); and (c) the metallic cylinder at another place
when the plane waves are incident vertically from the bottom to the
top.}

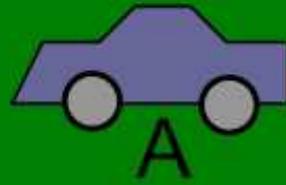
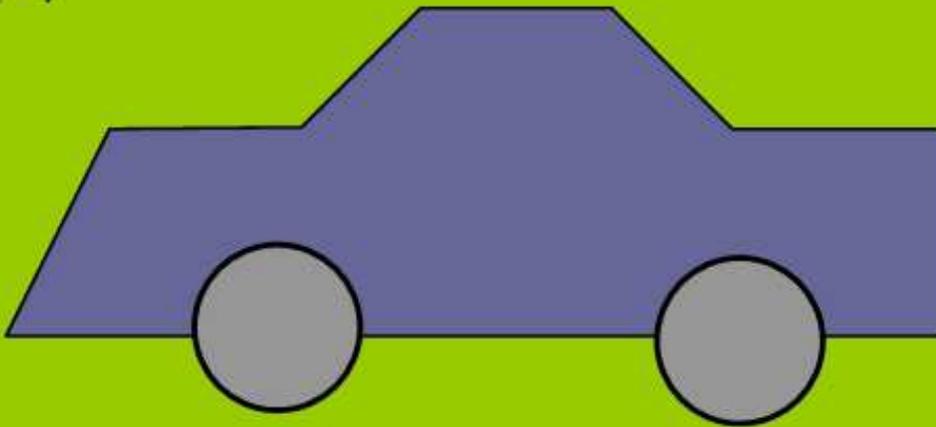
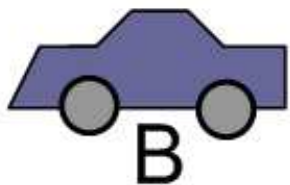
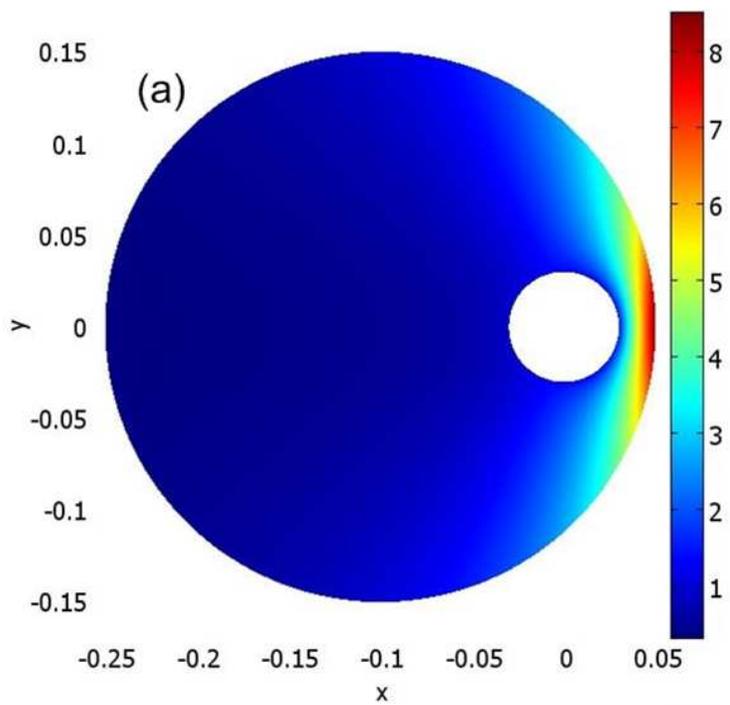
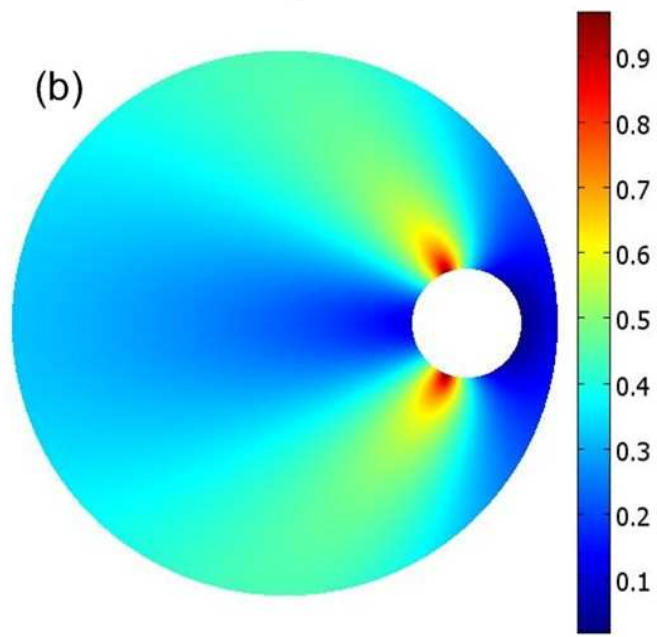
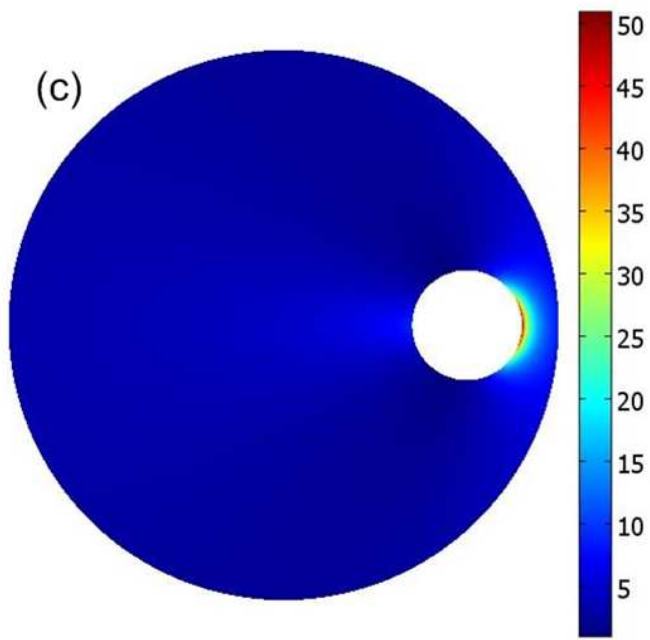
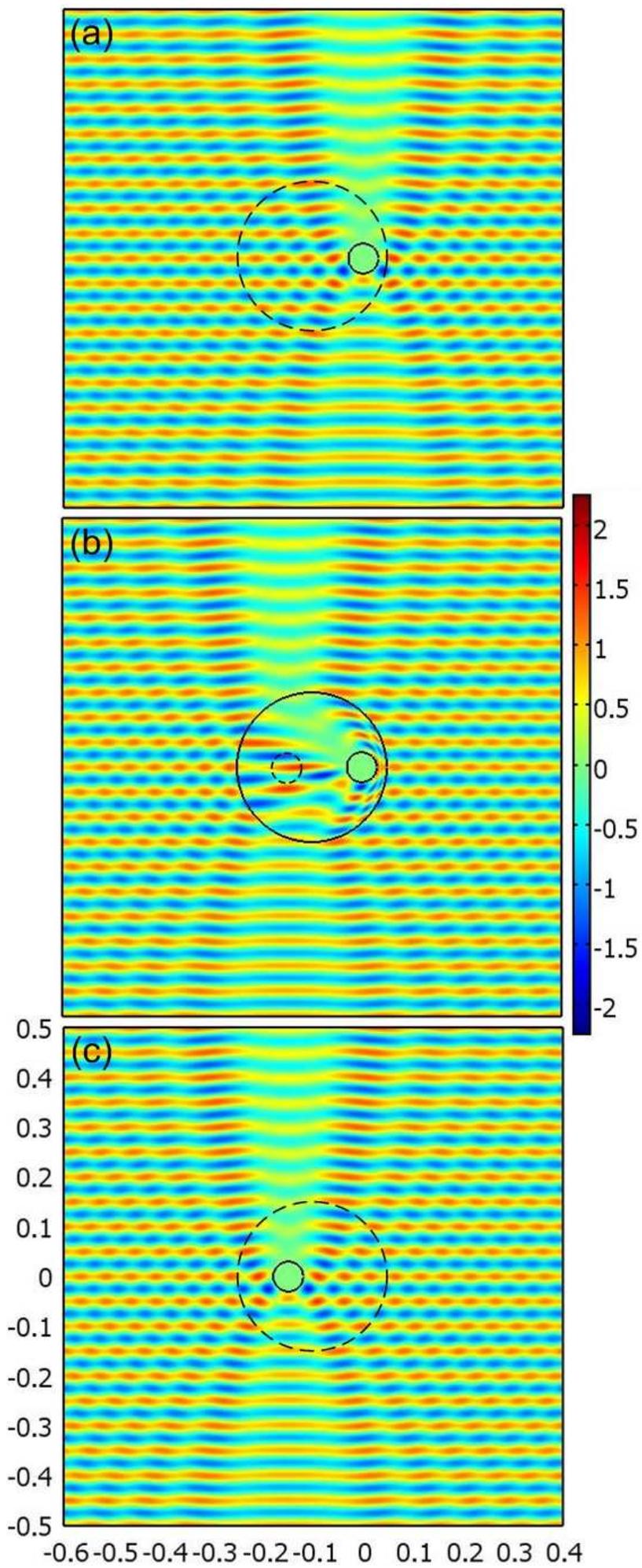
\end{document}